# Deterministic and non-volatile switching of all-van der Waals spin-orbit torque system above room temperature without external magnetic fields

Shivam N. Kajale[1], Thanh Nguyen[2], Mingda Li[2], Deblina Sarkar[1,*]

[1] MIT Media Lab, Massachusetts Institute of Technology, Cambridge, MA, 02139, USA
[2] Department of Nuclear Science and Engineering, Massachusetts Institute of Technology, Cambridge, MA, 02139, USA
[*] deblina@mit.edu

## Abstract

Two-dimensional van der Waals (vdW) magnetic materials hold promise for the development of high-density, energy-efficient spintronic devices for memory and computation. Recent breakthroughs in material discoveries and spin-orbit torque (SOT) control of vdW ferromagnets have opened a path for integration of vdW magnets in commercial spintronic devices. However, a solution for field-free electric control of perpendicular magnetic anisotropy (PMA) vdW magnets at room temperatures, essential for building compact and thermally stable spintronic devices, is still missing. Here, we report the first demonstration of field-free deterministic and non-volatile switching of a PMA vdW ferromagnet, $Fe_3GaTe_2$ above room temperature (up to 320 K). We use the unconventional out-of-plane anti-damping torque from an adjacent $WTe_2$ layer to enable such switching with a low current density of $2.23 \times 10^6$ A/cm$^2$. This study exemplifies the efficacy of low-symmetry vdW materials for spin-orbit torque control of vdW ferromagnets and provides an all-vdW solution for the next generation of scalable and energy-efficient spintronic devices.

## Introduction

The discovery of emergent magnetism in two-dimensional van der Waals (vdW) materials[1–3] has broadened the material space for developing spintronic devices for energy-efficient, non-volatile memory and computing applications[4–8]. These applications are particularly well-served by perpendicular magnetic anisotropy (PMA) ferromagnets, which allow fabrication of nanometer scale, high-density and thermally stable spintronic devices. vdW materials provide strong PMA alternatives[9–12] to the few bulk optimal material systems, like CoFeB/MgO[13–15], while providing key advantages like scalability down to monolayer thicknesses, and still maintaining an atomically smooth interface and minimal intermixing with the tunnel barrier of a magnetic tunnel junction (MTJ). Ability to switch the vdW PMA ferromagnets above room temperature is necessary for viable applications to harness these capabilities. Hence, recent reports on achieving current controlled switching of vdW PMA ferromagnets at room temperature are promising[16,17]. However, existing schemes for room temperature current control of vdW ferromagnets utilize spin-orbit torque (SOT) from heavy-metals or topological insulators and require application of an in-plane magnetic field to allow deterministic switching. This poses challenges to the development of high-density, thermally stable SOT-switching devices using vdW ferromagnets. A very recent work (unpublished) has attempted to use asymmetric growth of Pt on $Fe_3GaTe_2$ (single edge-coverage) to artificially break lateral symmetry for showing field-free switching of the vdW ferromagnet up to 300 K[18]. However, such a mechanism is inherently unscalable precluding wafer scale processing and in addition, robust non-volatile switching remains to be achieved. Thus, the critical challenge of field-free, deterministic, and non-volatile control of PMA magnetism in vdW materials above room temperature has remained unsolved.

Here, we report the first demonstration of deterministic and non-volatile switching of a PMA vdW ferromagnet above room temperature without any external magnetic fields. We achieved this by building a bilayer SOT system of room-temperature PMA vdW ferromagnet, $Fe_3GaTe_2$ (FGaT) with the low symmetry vdW material $WTe_2$ to harness the unconventional out-of-plane anti-damping torque for SOT switching (Fig. 1A). While several approaches to enabling field-free SOT switching of PMA magnetization are possible, including STT-assisted SOT switching[19], anisotropy tilting in the ferromagnet[20,21], artificially breaking lateral symmetry[22], and utilizing intrinsically low symmetry spin-orbit coupling layers[23–27], we have employed $WTe_2$ because it is particularly interesting for control of vdW magnets allowing creation of vdW heterostructures and ensuring pristine interfaces and no lattice strain. Charge current injection along the low symmetry $a$-axis of $WTe_2$ generates an unconventional, out-of-plane anti-damping SOT, $\tau_{AD}^{OOP}$, of the form $\hat{m} \times \hat{z} \times \hat{m}$ ($\hat{m}$, $\hat{z}$ are unit vectors along ferromagnet magnetization and $WTe_2$/ferromagnet interface)[28,29] and this torque can be utilized for field-free switching of PMA ferromagnets[24–27]. However, this mechanism has not been previously demonstrated for room-temperature field-free switching of vdW materials. Employing our FGaT/$WTe_2$ heterostructure devices, we demonstrate deterministic switching using a low current density of $2.23 \times 10^6$ A/cm$^2$ up to 320 K. We also show that such field-free deterministic switching is seen exclusively when the charge current is injected parallel to the low-symmetry axis of $WTe_2$, asserting the role of crystal symmetry in enabling such field-free switching of PMA magnetism.

## Results

Our heterostructure devices use exfoliated sheets of FGaT and WTe$_2$, with patterned electrical contacts, and hexagonal boron nitride (hBN) encapsulation for air-stability, as illustrated schematically through Fig. 1A. The heterostructures were assembled using dry viscoelastic transfer process[30] and electrodes were patterned using a combination of e-beam lithography and e-beam evaporation of Ti/Au (more details in Methods). The T$_d$-phase of WTe$_2$ used here belongs to the *Pmn2$_1$* space group. As shown in Fig. 1B, the crystal structure of WTe$_2$ is such that it preserves mirror symmetry about the *bc*-plane ($\sigma_{bc}$), while it breaks the mirror symmetry along the *ac*-plane ($\sigma_{ac}$), where *c* is the out-of-plane crystallographic axis. As a result, spin-orbit coupling induced spin-accumulation, and consequently the spin-orbit torque, in response to a current flowing along the *a*-axis and the *b*-axis varies significantly. These two cases are treated in detail in the following discussion, using two devices, D1 with FGaT (25.8 nm)/WTe$_2$ (21.6 nm) and D2 with FGaT (17.9 nm)/WTe$_2$ (23.8 nm). An optical image of the device D1 is shown in Fig. 1C, with the FGaT, WTe$_2$ and hBN flakes indicated. The crystallographic *a* and *b*-axes of the WTe$_2$ flakes were identified using polarized Raman spectroscopy in the backscattering geometry $Z(\hat{\phi}\hat{\phi})\bar{Z}$, where $\hat{\phi}$ is a unit vector in the sample plane, along the azimuthal angle $\phi$ as defined in Fig. 1C. Fig. 1D shows a color plot of the polarized Raman spectra of the WTe$_2$ flake in D1 (see Supplementary Fig. S2 for D2). WTe$_2$ exhibits two types of prominent A$_g$ peaks with two-fold symmetries, which can be used to identify its *a* and *b*-axes[31,32]. The minima in the type-I peaks (81 cm$^{-1}$ and 212 cm$^{-1}$), which coincides with the maxima in the type-II peak (165 cm$^{-1}$) corresponds to the *a*-axis of the WTe$_2$ crystal.

Magneto-transport characterization of the FGaT/WTe$_2$ devices using anomalous Hall effect helps to establish that the inherent ferromagnetic characteristics of FGaT are well preserved in the heterostructure device and can be effectively probed through transverse voltage monitoring for current-induced magnetization switching experiments. Fig. 2A, B show the anomalous Hall effect curves for the device D1, for field swept along sample normal ($H \parallel c$) and temperatures in the range 10 K to 340 K. The device exhibits a large coercivity (up to 8.25 kOe at 10 K) at low temperatures, which diminishes with temperature (Fig. 2C) such that $H_c = 210$ Oe at 300 K and near-zero starting 330 K. The anomalous Hall resistance, $R_{xy}^{AHE}$ goes to zero above 320 K too, marking a ferromagnet to paramagnet transition between 320 K to 330 K. The anomalous Hall effect curve corresponding to field swept close to sample plane ($H \perp c$) is shown in Fig. 2D. It exhibits the characteristics of a PMA magnet, going to near-zero resistance values only at high in-plane magnetic fields, with an anisotropy field of about 35 kOe, corroborating that the strong perpendicular magnetic anisotropy of FGaT is preserved in the heterostructure device.

Fig. 3A provides a schematic representation of the spin-orbit torque mechanism at play when the applied current is parallel to the high-symmetry, *b*-axis. In this case, the applied current has no effect on the crystal's *bc*-mirror plane symmetry ($\sigma_{bc}$). In accordance with Curie's principle[33], since the causalities (crystal structure and applied current) preserve $\sigma_{bc}$, the resultant spin-current (and accumulation) must also preserve $\sigma_{bc}$. This forbids a vertical spin-polarization ($\sigma_z$) component in the vertically flowing spin-current, since the $\sigma_z$ pseudovector transforms anti-symmetrically upon reflection in the *bc*-plane. As a result, the spin-accumulation at the FGaT/WTe$_2$ interface only has an in-plane spin-polarization, similar to the case of heavy metal/ferromagnet and topological insulator/ferromagnet systems. Such an in-plane spin accumulation can only produce deterministic switching in the presence of an externally applied field along the current direction. Fig. 3B shows the response of device D1 to current pulses applied along the *b*-axis of WTe$_2$ in the absence of any external field, at 300 K. As expected, the in-plane anti-damping torque from spin-accumulation at the FGaT/WTe$_2$ interface drives the FGaT magnetization in-plane ($m_z = 0$) resulting in a near-zero anomalous Hall resistance, for a current magnitude of about ±4.5 mA (9.51 × 10$^5$ A/cm$^2$). Upon lowering the current drive to zero, the FGaT remains effectively demagnetized as its various domains orient randomly due to lack of a symmetry breaking field. The four curves in Fig. 3B verify this for all combinations of current drive (positive or negative) and initial magnetization direction ($m_z = \pm 1 \equiv R_{xy} = \pm 1.2\ \Omega$). The initial magnetization state is set by applying a field of ±2 kOe along the sample normal before starting current sweeps. Contrary to the above case, driving a current of the same magnitude in the presence of a non-zero external field ($H = \pm 500$ Oe) parallel to current axis, $H \parallel I \parallel b$, results in deterministic, partial switching of the FGaT magnetization. As shown in Fig. 3C, reversing the direction of applied field reverses the chirality of the current-induced switching loops, as is expected for such a system. Field-assisted deterministic and non-volatile switching of out-of-plane magnetization of FGaT could also be achieved in D1 for the case of $H \parallel I \parallel b$ (Fig. 3D) at 300 K.

In contrast to the above discussed case, when current is applied along the low-symmetry *a*-axis of WTe$_2$, the applied current breaks the *bc*-mirror plane symmetry ($\sigma_{bc}$). Thus, the causalities break both the mirror plane symmetries ($\sigma_{ac}$ broken by crystal structure, $\sigma_{bc}$ broken by applied current), and a vertical spin-polarization component in the vertical spin-current is now permissible. This scenario is depicted schematically in Fig. 4B. The vertical component of spin-accumulation at the FGaT/WTe$_2$ interface can now apply a symmetry breaking, unconventional, out-of-plane anti-damping spin orbit torque, $\tau_{AD}^{OOP}$, on the FGaT magnetization. $\tau_{AD}^{OOP}$ is anti-symmetric in current and hence, the FGaT magnetization can be toggled deterministically between $m_z = \pm 1$ by applying positive and negative current pulses. Device D2, with current applied along the *a*-axis of its WTe$_2$ flake is used to study this scenario. Details on the device, its Raman spectra and magneto-transport data are included in Supplementary Fig. S2 and S3. Fig. 4A shows the field-free current induced switching loops of D2 for temperatures ranging from 300 K to 325 K. At 300 K, complete switching could be induced using ±8 mA (see Supplementary Fig. S4), equivalent to a current density of 2.23 × 10$^6$ A/cm$^2$. Increasing the temperature from 300 K to 325 K resulted in shrinking of the anomalous Hall resistance splitting, until no clear looping behavior could be observed at 330 K and beyond (Fig. 4C). This aligns with the fact that magnetization of FGaT would decrease with increasing temperature, resulting in a decreasing $R_{xy}^{AHE}$ until it eventually vanishes beyond its Curie temperature (320 K – 330 K). Fig. 4D shows the field-free deterministic and non-volatile switching of PMA magnetization of FGaT by a train of current pulses, 1 ms long and ±8 mA in amplitude, applied along the low-symmetry axis of WTe$_2$, $I \parallel a$, at 300 K. We could observe such deterministic switching right up to 320 K as reported in Supplementary Fig. S6, providing the first demonstration of field-free, deterministic switching of out-of-plane magnetization in a vdW ferromagnet above room temperature.

## Conclusion

We utilize the unconventional, out-of-plane anti-damping spin orbit torque, $\tau_{AD}^{OOP}$, generated from WTe$_2$ upon charge current injection along its low-symmetry *a*-axis to switch the magnetization of underlying FGaT, in the FGaT/WTe$_2$ heterostructure devices. We clearly show that the $\tau_{AD}^{OOP}$ induced field-free switching occurs exclusively for charge current injection along WTe$_2$ *a*-axis, while charge injection along the *b*-axis results in demagnetization of underlying FGaT. Thus, we have reported the first demonstration of field-free magnetization switching of a perpendicular magnetic anisotropy vdW ferromagnet above room temperature (up to 320 K) using a low current density of 2.23 × 10$^6$ A/cm$^2$. The proposed all-vdW architecture can also provide unique advantages like improved interface quality needed for efficient spin-orbit torques, possibilities for gate-voltage tuning to assist SOT switching, and prospects for flexible and transparent spintronic technologies. This work asserts the role of crystal symmetry in spin-orbit coupling layers of an SOT switching device using a low-symmetry vdW material, and provides a new, scalable all-vdW approach to developing energy-efficient spintronic devices.

## Methods

### Device fabrication

The Fe$_3$GaTe$_2$/WTe$_2$ devices reported here were fabricated using heterostructure assembly of exfoliated vdW flakes. Bulk FGaT was grown using a previously reported process[17]. Bulk WTe$_2$ and hBN were commercially sourced from HQ Graphene and Ossila, respectively. FGaT flakes were exfoliated on Si/SiO$_2$ (280 nm) substrates using mechanical exfoliation. WTe$_2$ flakes, exfoliated on PDMS stamps were transferred on to selected FGaT flakes using the dry viscoelastic transfer process. Electrodes were then patterned on the FGaT/WTe$_2$ heterostructure using a combination of e-beam lithography with the positive e-beam resist PMMA 950, and e-beam evaporation of Ti/Au (5 nm/60 nm). The devices were then encapsulated with thick exfoliated flakes of hBN, using dry viscoelastic transfer. All exfoliation and vdW transfer processes were performed inside the inert environment of a N$_2$-filled glovebox (O$_2$, H$_2$O < 0.01 ppm). Thicknesses of the constituent flakes were characterized after encapsulation using a Cypher VRS AFM. Polarized Raman spectra of WTe$_2$ flakes was acquired using a 532 nm laser with a WITec Alpha300 Apyron Confocal Raman microscope, by rotating the polarizer and analyzer while the sample was static.

### Transport measurements

All transport measurements were performed in a 9 T PPMS DynaCool system. Measurements were performed by sourcing current using a Keithley 6221 current source and measuring the transverse voltage across the devices, using a Keithley 2182A nanovoltmeter. Anomalous Hall effect measurements with field sweeps were performed using a drive current of 50 – 200 µA. For the current-induced switching measurements, a 1 ms pulse of write-current was followed by 999 ms of read pulses (±200 µA). Field could be applied in and out of the sample plane using the PPMS' horizontal rotator module.

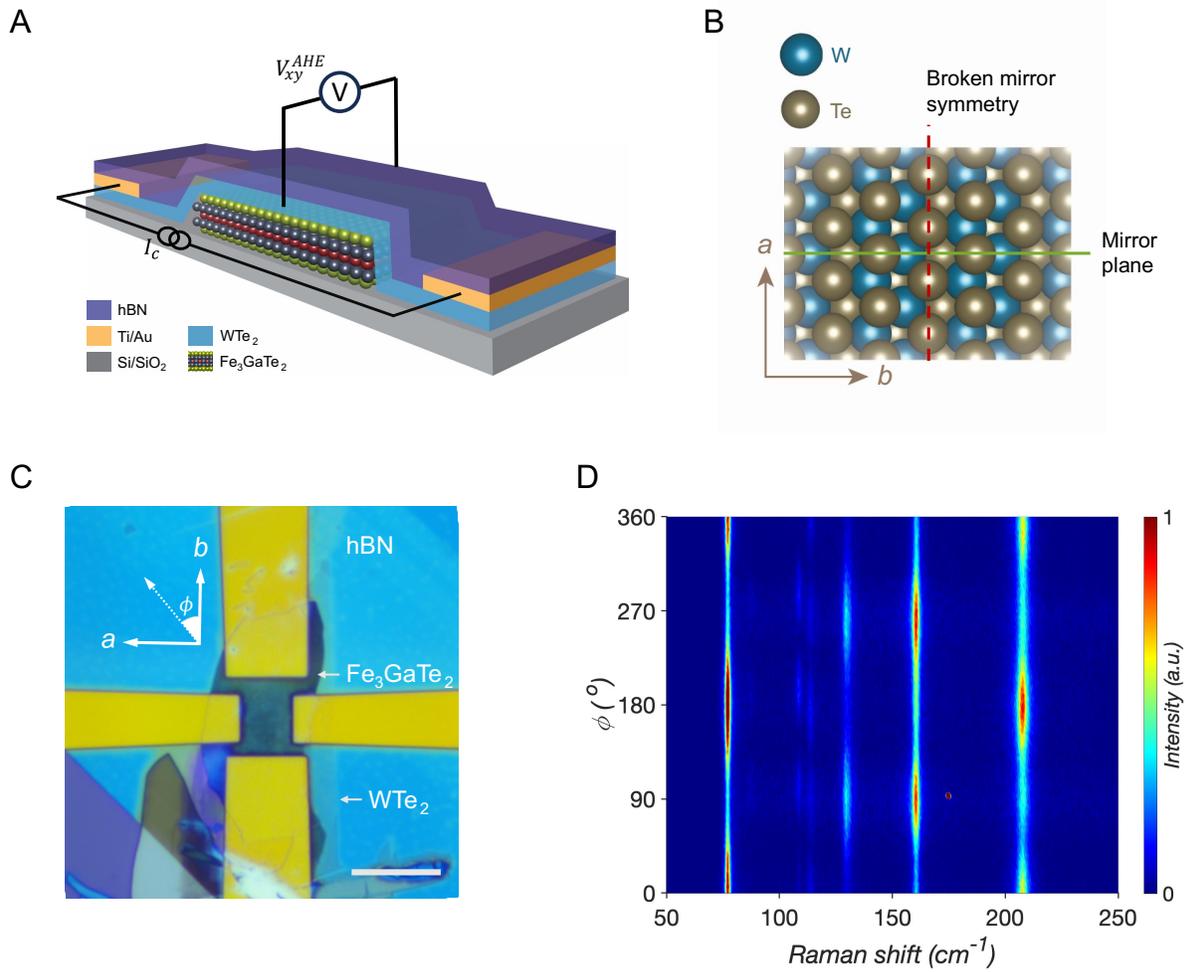

**Fig. 1: Fe₃GaTe₂/WTe₂ heterostructure device.** **(A)** Schematic diagram of the Fe₃GaTe₂/WTe₂ heterostructure devices used in this study. **(B)** Schematic model of WTe₂ crystal's *ab*-plane, with the *a* and *b*-axes labelled. The crystal preserves mirror-plane symmetry in the *bc*-plane while breaks it in the *ac*-plane. **(C)** Optical image of device D1, with the WTe₂ (21.6 nm), FGaT (25.8 nm) and hBN flakes labelled. Crystallographic axes of the WTe₂ flake (determined through polar Raman spectra) and the definition of azimuthal angle $\phi$ in the Raman spectra are also indicated. Scale bar: 10 μm **(D)** Polarized Raman spectra of the WTe₂-flake in (C). The minima (*maxima*) in type-I $A_g$ modes at 81 cm$^{-1}$ and 212 cm$^{-1}$ (*type-II $A_g$ mode at 165 cm$^{-1}$*) around $\phi = 90°$ corresponds to the *a*-axis of the WTe₂ flake.

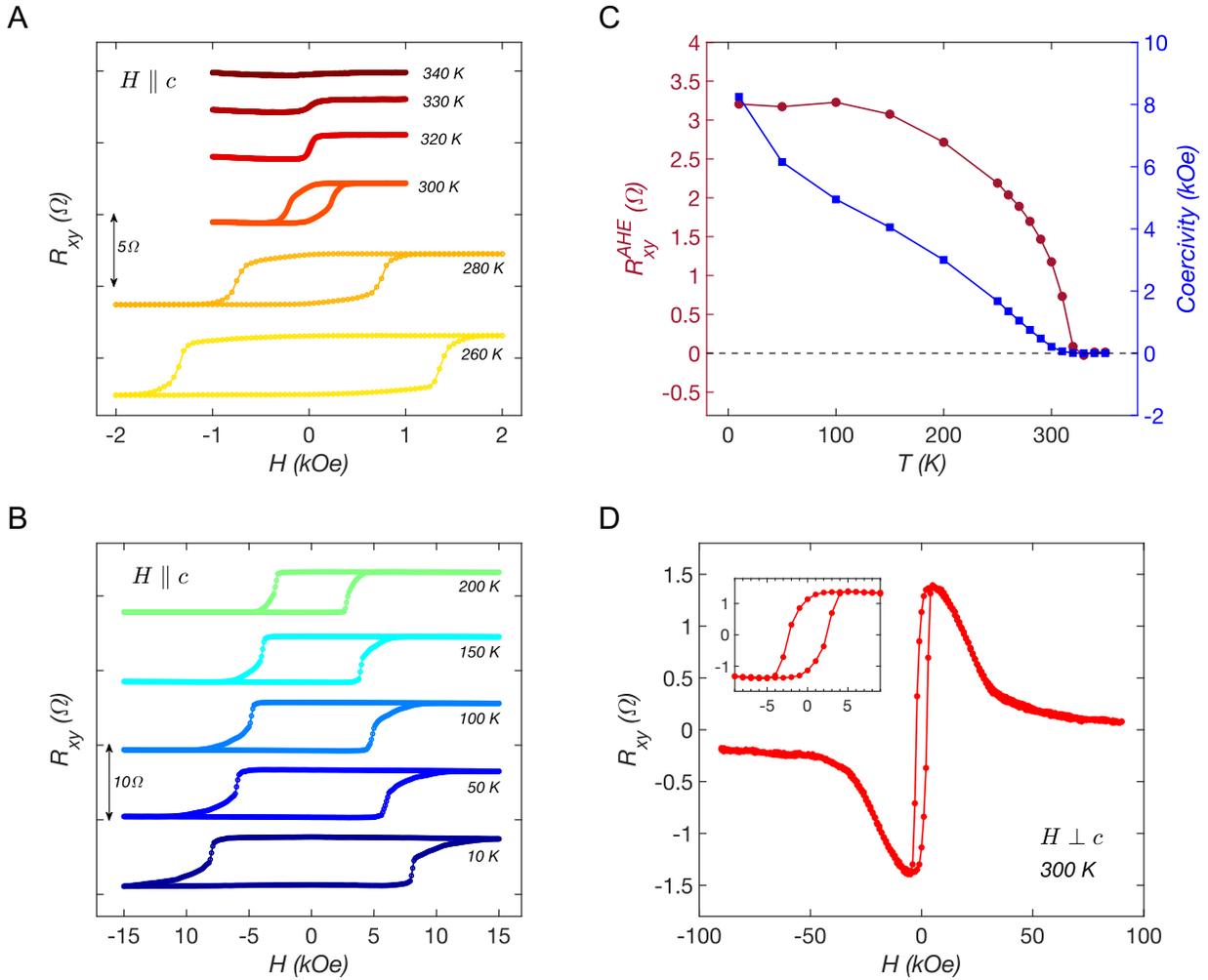

**Fig. 2:** <u>Magneto-transport characterization of FGaT/WTe$_2$ devices.</u> **(A, B)** Anomalous Hall effect measurements for field swept out of the sample plane ($H \parallel c$) at varying temperatures up to 340 K. Data presented corresponds to device D1 (Fig. 1C). Data is offset along y-axis for clarity. **(C)** Variation of anomalous Hall resistance ($R_{xy}^{AHE}$, left y-axis) and coercivity ($H_c$, right y-axis) with temperature. **(D)** Anomalous Hall effect measurement for field swept close to sample plane ($H \perp c$), with transverse resistance reaching near-zero level at high fields, indicative of the strong perpendicular magnetic anisotropy of FGaT being preserved in the FGaT/WTe$_2$ heterostructure devices. Inset: Zoomed in view of the AHE curve for low fields.

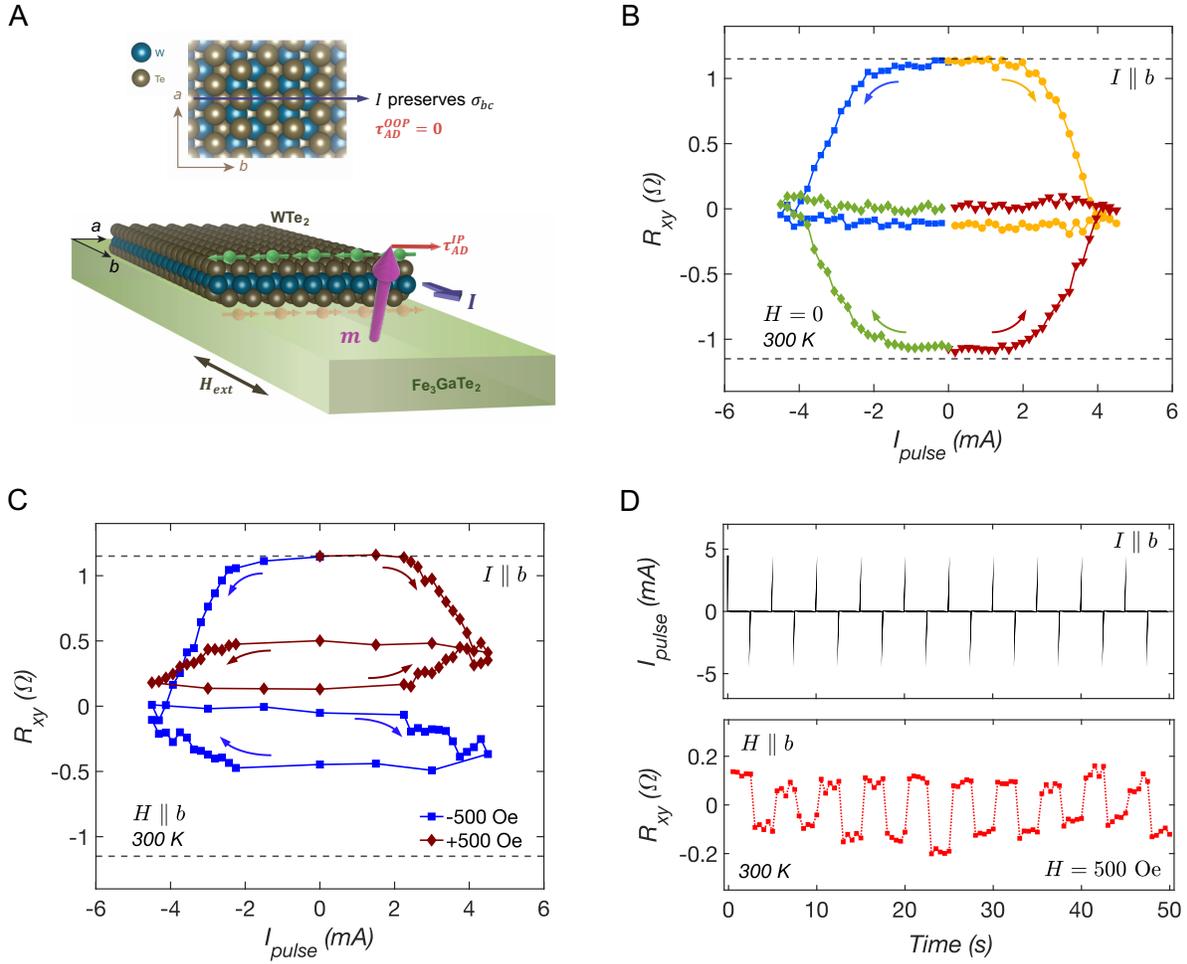

**Fig. 3: Field-assisted (only) switching for $I \parallel b$.** **(A)** Schematic illustration of the scenario where current is sourced along the high-symmetry axis, $I \parallel b$. Symmetry constraints allow only an in-plane component of spin-accumulation along the FGaT/WTe$_2$ interface, resulting in a non-zero in-plane anti-damping torque ($\tau_{AD}^{IP} \neq 0$) but a zero out-of-plane anti-damping torque ($\tau_{AD}^{OOP} = 0$). **(B)** Response of the device to current pulses applied along the *b*-axis for zero external field at 300 K. The blue and green (*yellow and red*) curves correspond to current pulses swept from $0 \rightarrow -4.5$ mA $\rightarrow 0$ ($0 \rightarrow +4.5$ mA $\rightarrow 0$ mA), for the device initialized at $m_z = 1$ and $m_z = -1$, respectively. The device undergoes complete demagnetization by 4.5 mA in all the four cases. **(C)** Current sweeps up to 4.5 mA result in partial magnetization switching in the presence of an externally applied field, $H \parallel b \parallel I$ of $\pm 500$ Oe, with changing the direction of field resulting in chirality reversal of the current-induced switching curves. Black dashed lines in (B) and (C) correspond to $m_z = \pm 1$. **(D)** Field-assisted deterministic, non-volatile switching of FGaT magnetization using a train of 1 ms long current pulses, $\pm 4.5$ mA in magnitude, under $+500$ Oe in-plane magnetic field, $H \parallel b \parallel I$.

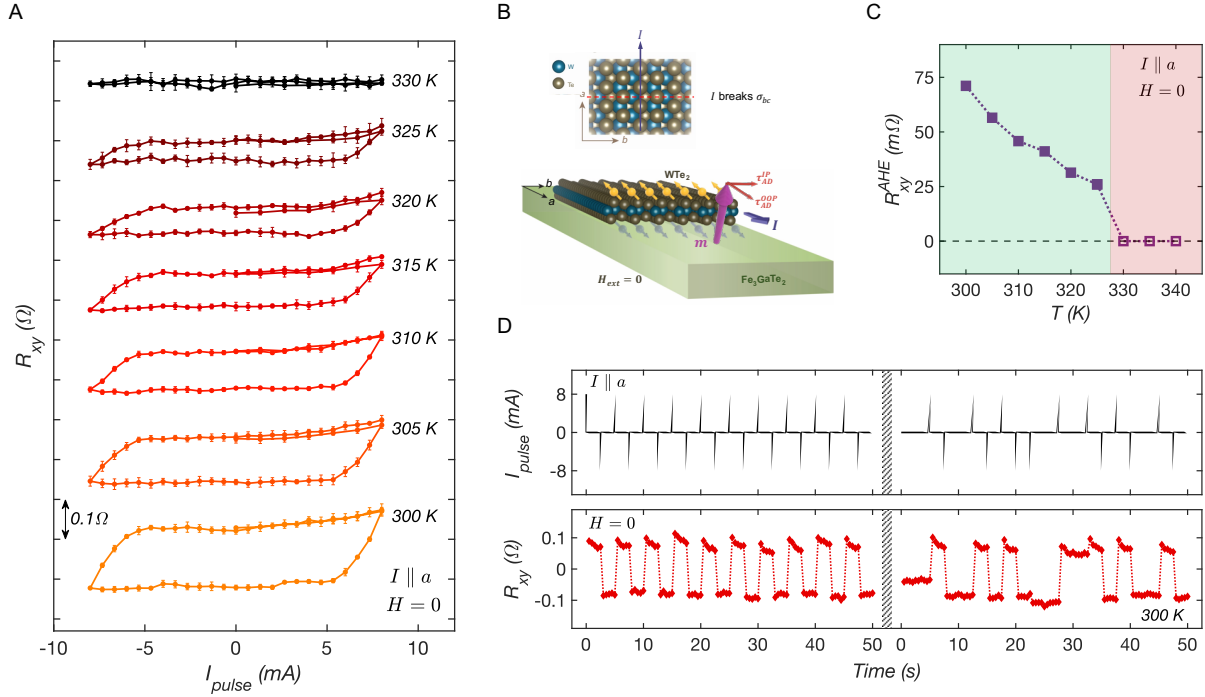

**Fig. 4:** <u>Field-free switching for $I \parallel a$.</u> **(A)** Response of the device D2 to current pulse sweeps, along $a$-axis, for varying temperatures without any external field. The curve at each temperature is an average of four consecutive current pulse sweeps acquired for that temperature, with error bars indicating standard deviation of each data point across the four sweeps (Individual sweeps reported in Supplementary Fig. S5). Data offset along y-axis for clarity. **(B)** Schematic illustration of this scenario where current is sourced along the low-symmetry axis, $I \parallel a$. Broken mirror plane symmetries allow an out-of-plane component of spin-accumulation along the FGaT/WTe$_2$ interface, resulting in a non-zero out-of-plane anti-damping torque ($\tau_{AD}^{IP} \neq 0$), asymmetric in current direction, enabling field-free deterministic switching of the underlying FGaT's magnetization. **(C)** Temperature dependence of the anomalous Hall resistance splitting in the current-induced switching loops. Clear switching can be observed up to 325 K (green region), with decreasing $R_{xy}^{AHE}$ denoted with solid square point, while no clear switching loops could be observed starting 330 K (red region), and hence the $R_{xy}^{AHE}$ is set to zero (hollow square points). **(D)** Demonstration of field-free, deterministic, non-volatile switching of out-of-plane FGaT magnetization in the FGaT/WTe$_2$ device using 1 ms long pulses of current, ±8 mA in amplitude, applied along the $a$-axis. The data is acquired at 300 K in two sets of 50 s long pulsing sequences, with periodic and randomized current pulses, respectively.

# Deterministic and non-volatile switching of all-van der Waals spin-orbit torque system above room temperature without external magnetic fields

Shivam N. Kajale[1], Thanh Nguyen[2], Mingda Li[2], Deblina Sarkar[1,*]

[1] MIT Media Lab, Massachusetts Institute of Technology, Cambridge, MA, 02139, USA
[2] Department of Nuclear Science and Engineering, Massachusetts Institute of Technology, Cambridge, MA, 02139, USA
[*]deblina@mit.edu

Fig. S1: Topographical data for device D1.

Fig. S2: Device D2, its Raman spectra and topography.

Fig. S3: Magneto-transport characterization of D2.

Fig. S4: Effect of increasing peak current in current sweeps in D2.

Fig. S5: Current-pulsing loops across varying temperatures.

Fig. S6: Deterministic switching up to 320 K.

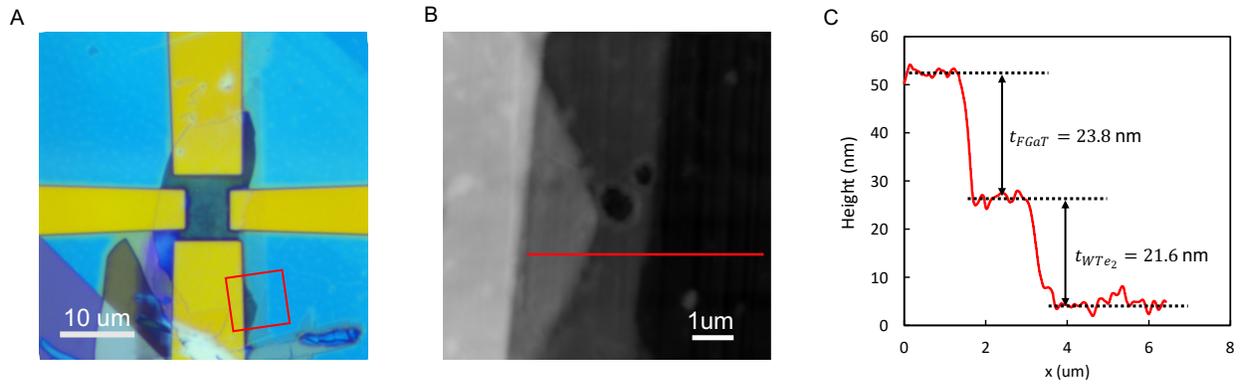

**Fig. S1:** <u>Topographical data for device D1.</u> **(A)** Optical image of the device, with red box indicating the region used for AFM measurements. **(B)** AFM topography micrograph of the region in red box. **(C)** Height profile along the red line in (B).

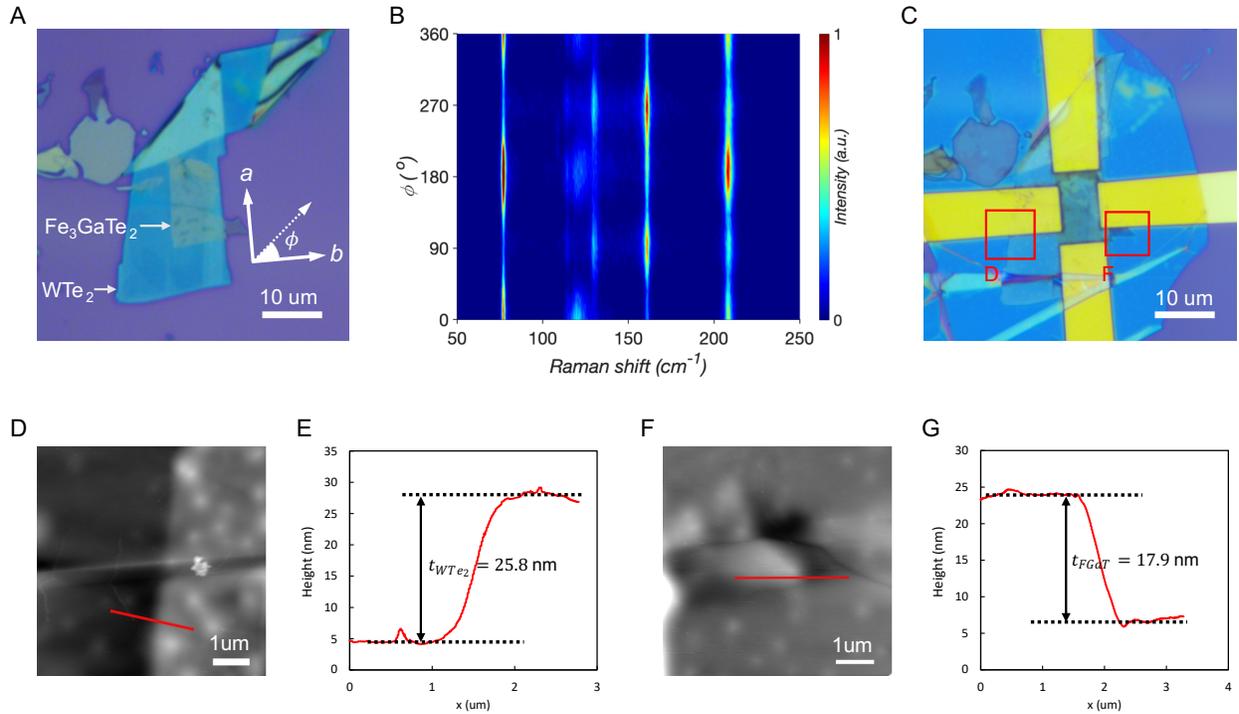

**Fig. S2:** <u>Device D2, its Raman spectra and topography.</u> **(A)** Optical image of the FGaT/WTe2 heterostructure before patterning electrodes. The crystallographic axes of WTe$_2$ (determined using polar Raman spectra) and the definition of azimuthal angle $\phi$ in the polar Raman measurements is indicated. **(B)** Polarized Raman spectra of the WTe$_2$ flake in D2. **(C)** Optical image of the device D2, after patterning electrodes and encapsulation with hBN. Red boxes correspond to area scanned in AFM for determining the thicknesses of the constituent WTe$_2$ (box D) and FGaT (box F) flakes. **(D)** AFM topography micrograph of red box D and **(E)** the height profile along the red line. **(F)** AFM topography micrograph of the red box F and **(G)** the height profile along the red line.

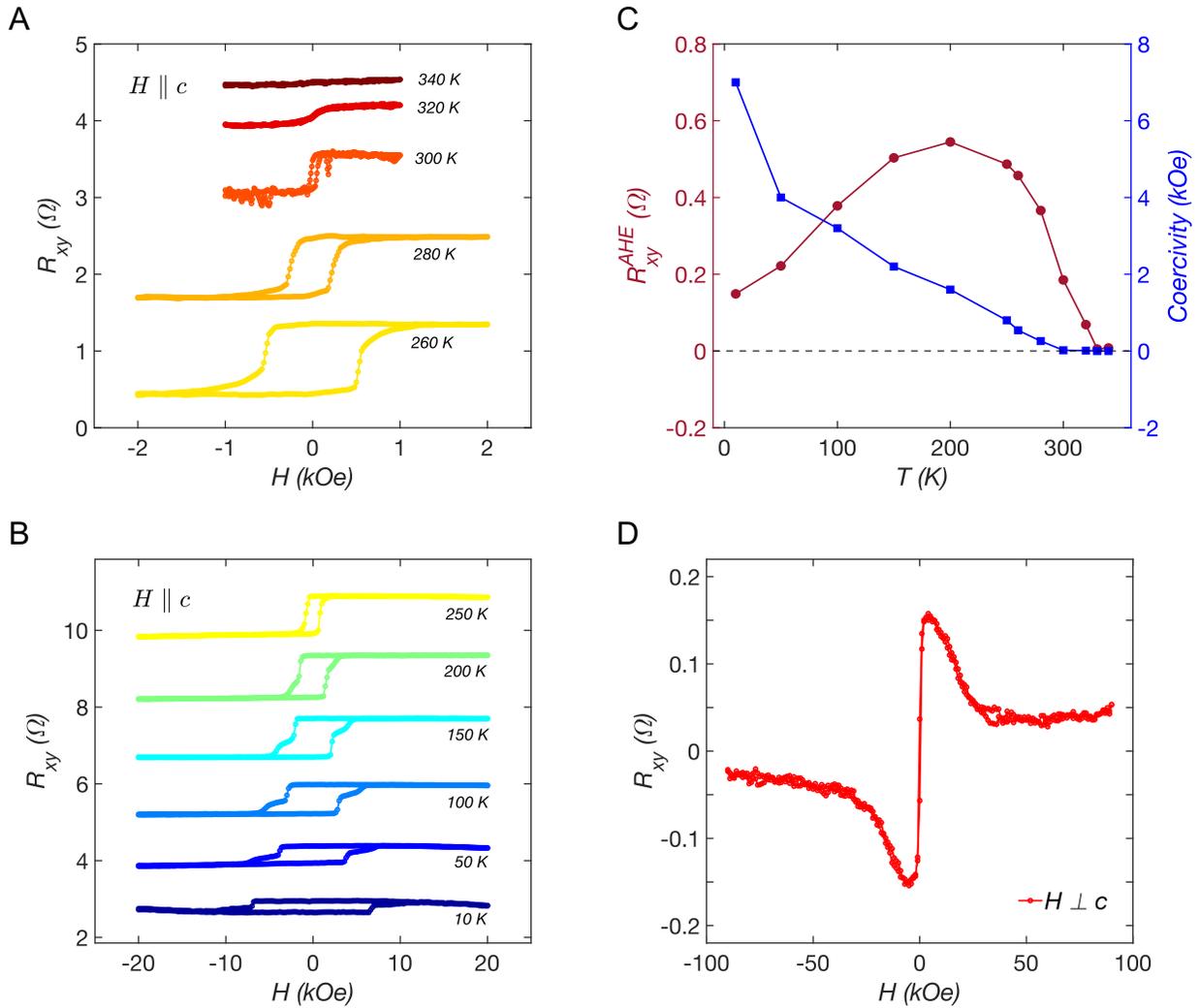

**Fig. S3:** Magneto-transport characterization of D2. **(A, B)** Anomalous Hall effect measurements for field swept out of the sample plane ($H \parallel c$) for varying temperatures up to 340 K. Data is offset along y-axis for clarity. **(C)** Variation of anomalous Hall resistance ($R_{xy}^{AHE}$, left y-axis) and coercivity ($H_c$, right y-axis) with temperature. **(D)** Anomalous Hall effect measurement for field swept close to sample plane ($H \perp c$), indicative of the strong perpendicular magnetic anisotropy of FGaT being preserved in the FGaT/WTe$_2$ heterostructure device.

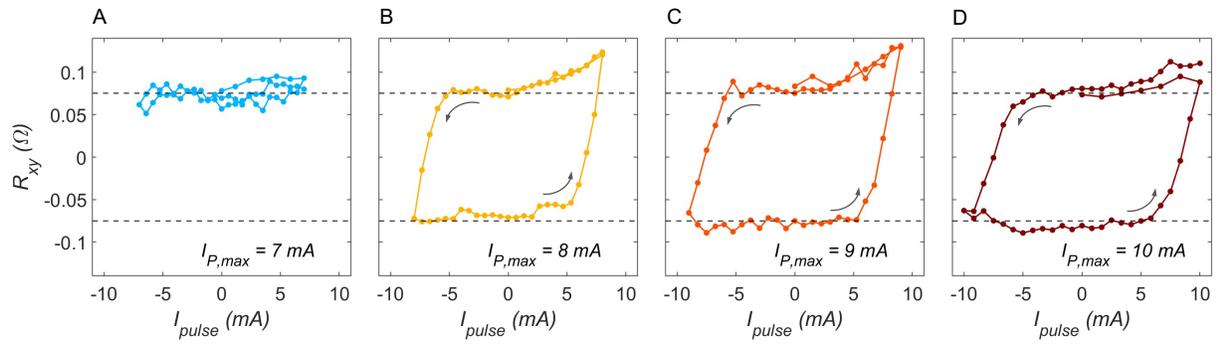

**Fig. S4:** <u>Effect of increasing peak current in current sweeps in D2.</u> **(A)** No clear switching behavior observed when current sweep is limited to a maximum pulse amplitude $I_{P,max}$ of $\pm 7$ mA. **(B)** A clear, cyclic switching curve is observed on increasing $I_{P,max}$ to $\pm 8$ mA. Further increasing $I_{P,max}$ to **(C)** $\pm 9$ mA and **(D)** $\pm 10$ mA does not increase the loop's vertical splitting notably. Thus, switching is deemed to be near-complete by 8mA. Black dashed lines are a visual aid denoting the same loop splitting.

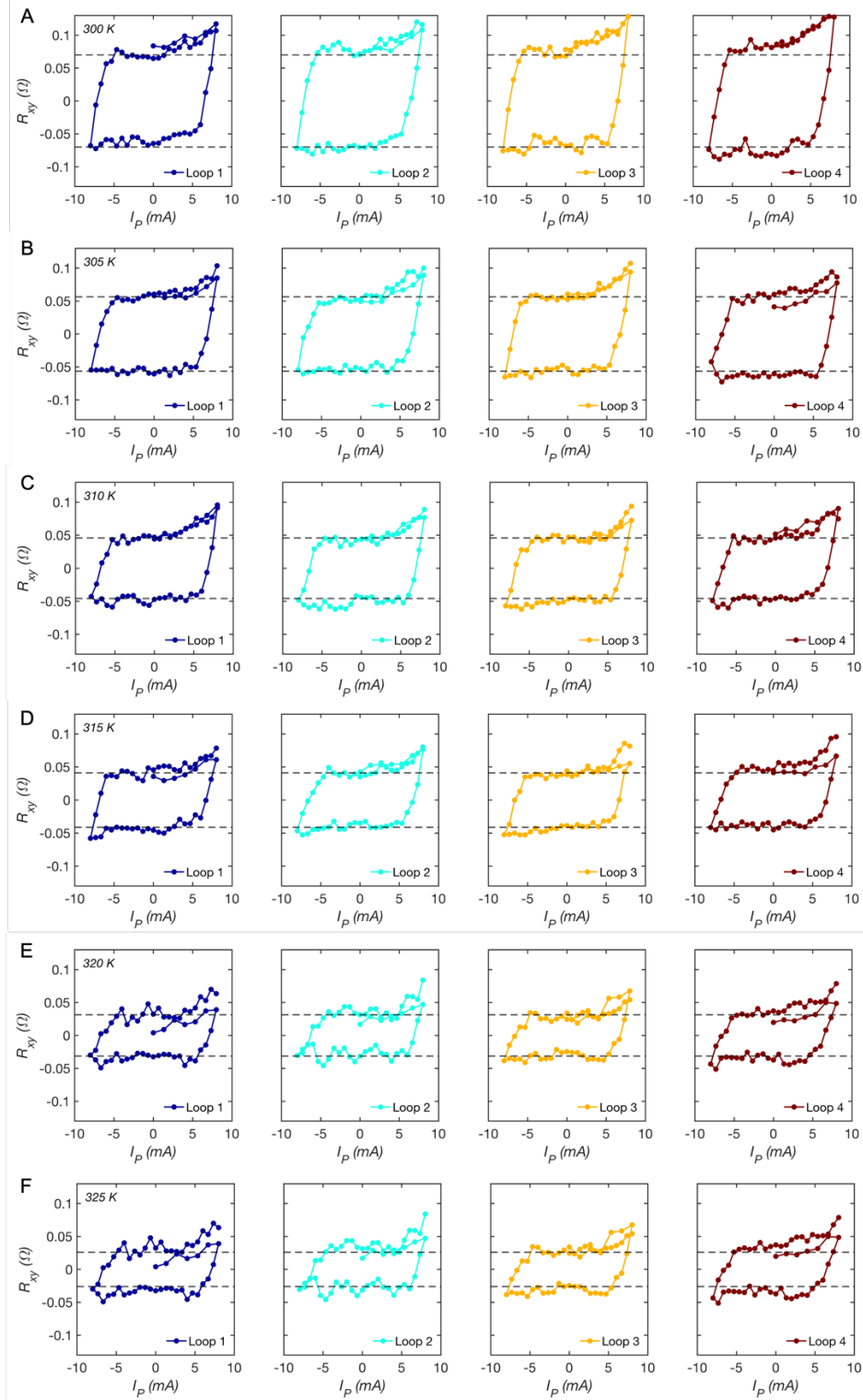

**Fig. S5:** <u>Current-pulsing loops across varying temperatures.</u> **(A)** Four consecutive current pulsing loops acquired for D2, with $I \parallel a$, without any external field (no field-assisted initialization between consecutive loops either) at 300 K. Black dashed lines are a visual aid denoting the same loop splitting. **(B-F)** Similar data for temperatures 305 K – 325 K in steps of 5 K.

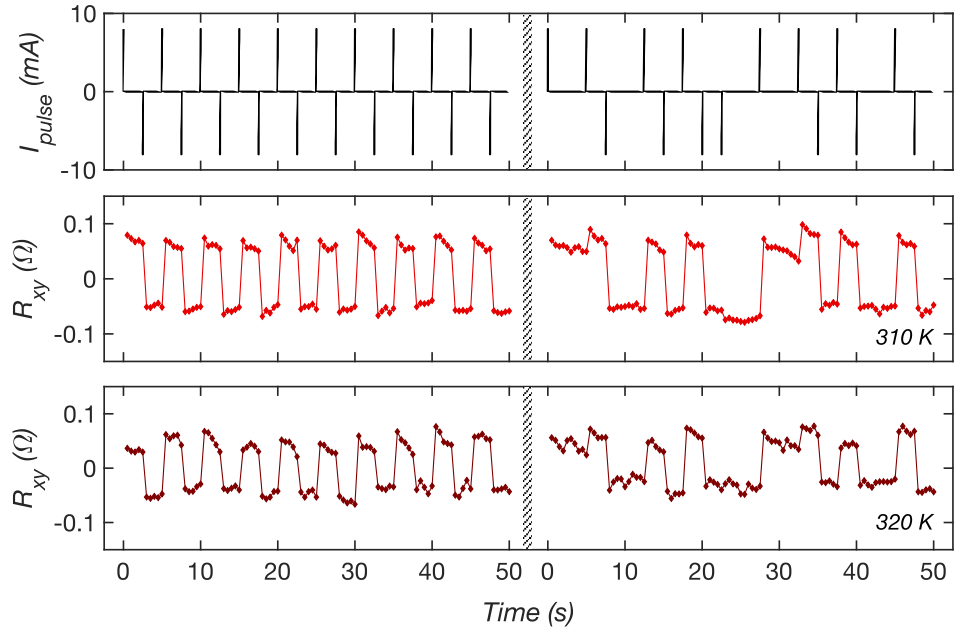

**Fig. S6:** <u>Deterministic switching up to 320 K.</u> Field free deterministic, non-volatile switching of the OOP magnetization of FGaT in device D2, using the train of current pulses, 1 ms long and $\pm 8$ mA in magnitude (top panel), with $I \parallel a$, at 310 K (middle panel) and 320 K (lower panel). The data is acquired in two sets of 50 s long pulsing sequences, with periodic and randomized current pulses, respectively.